\documentclass[usenatbib]{mn2e}

\bibpunct{(}{)}{;}{a}{}{,}

\newcommand{\ltsima}{$\buildrel < \over \sim$}
\newcommand{\lsim}{\lower.5ex\hbox{\ltsima}}
\newcommand{\gtsima}{$\buildrel > \over \sim$}
\newcommand{\gsim}{\lower.5ex\hbox{\gtsima}}
\newcommand{\xmm}{XMM-Newton}

\newcommand{\xmma}{\hbox{XMMU\,J010633.1-731543}}
\newcommand{\xmmb}{\hbox{XMMU\,J010743.1-715953}}

\usepackage{graphicx}
\usepackage{amssymb}
\usepackage{epsfig}
\usepackage{natbib}
\def\ion#1#2{#1$\;${\small\rm\@Roman{#2}}\relax}

\title[Two new X-ray sources in the SMC]{The XMM-Newton survey of the Small Magellanic Cloud:\\
       \xmma and \xmmb, two new Be/X-ray binary systems.
       \thanks{Based on observations with
               XMM-Newton, an ESA Science Mission with instruments and contributions
               directly funded by ESA Member states and the USA (NASA)}}

\author[M.J. Coe et al.]{M. J.~Coe$^{1}$, F. Haberl$^{2}$, R. Sturm$^{2}$, E. S. Bartlett$^{1}$, \and D. Hatzidimitriou$^{3}$, L.J. Townsend$^{1}$, A. Udalski$^{4}$, S. Mereghetti${^5}$ \& M. Filipovi{\'c}$^{6}$\\
$^{1}$ School of Physics and Astronomy, University of Southampton, SO17
1BJ, UK\\
$^{2}$ Max-Planck-Institut f\"ur extraterrestrische Physik,
           Giessenbachstra{\ss}e, 85748 Garching, Germany\\
$^{3}$ Department of Astrophysics, Astronomy and Mechanics, Faculty of Physics, University of Athens, Panepistimiopolis, \\ 15784 Zografos, Athens, Greece; IESL, Foundation for Research and Technology, 71110, Heraklion, Greece \\
$^{4}$ Warsaw University Observatory, Aleje Ujazdowskie 4, 00-478 Warsaw, Poland \\
$^{5}$ INAF, Istituto di Astrofisica Spaziale e Fisica Cosmica Milano, via E. Bassini 15, 20133 Milano, Italy \\
$^{6}$ University of Western Sydney, Locked Bag 1797, Penrith South DC, NSW1797, Australia
}

\begin{document}

\date{26 April 2012}

\pagerange{\pageref{firstpage}--\pageref{lastpage}} \pubyear{2002}

\maketitle

\label{firstpage}

\begin{abstract}

{   In the course of the XMM-Newton survey of the Small Magellanic Cloud (SMC), two new bright X-ray sources were discovered exhibiting the spectral characteristics of High Mass X-ray Binaries - but revealing only weak evidence for pulsations in just one of the objects(at 153s in \xmmb).
 The accurate X-ray source locations permit the identification of these X-ray source with Be stars, thereby strongly suggesting these systems are new Be/X-ray binaries. From blue spectra the proposed classification for \xmma ~is B0.5-1Ve and for \xmmb ~it is B2IV-Ve.

}

\end{abstract}


\section{Introduction and background}

The Be/X-ray systems represent the largest sub-class of all High Mass X-ray Binaries (HMXB).  A survey of the literature reveals that of the $\sim$240 HMXBs known in our Galaxy and the Magellanic Clouds (Liu et al., 2005, 2006), $\ge$50\%
fall within this class of binary.  In fact, in recent years it has emerged that there is a substantial population of HMXBs in the SMC comparable in number to the Galactic population. Though unlike the Galactic population, all except one of the SMC HMXBs are Be star systems.  In these systems the orbit of the Be star
and the compact object, presumably a neutron star, is generally wide
and eccentric.  X-ray outbursts are normally associated with the
passage of the neutron star close to the circumstellar disk (Okazaki
\& Negueruela 2001), and generally are classified as Types I or II (Stella, White \& Rosner, 1986). The Type I outbursts occur periodically at the time of the periastron passage of the neutron star, whereas Type II outbursts are much more extensive and occur when the circumstellar material expands to fill most, or all of the orbit. General reviews of such HMXB systems may
be found in Negueruela (1998), Corbet et al. (2009) and Coe et al. (2000, 2009).

\begin{figure*}
  \resizebox{\hsize}{!}{\includegraphics[angle=0,clip=]{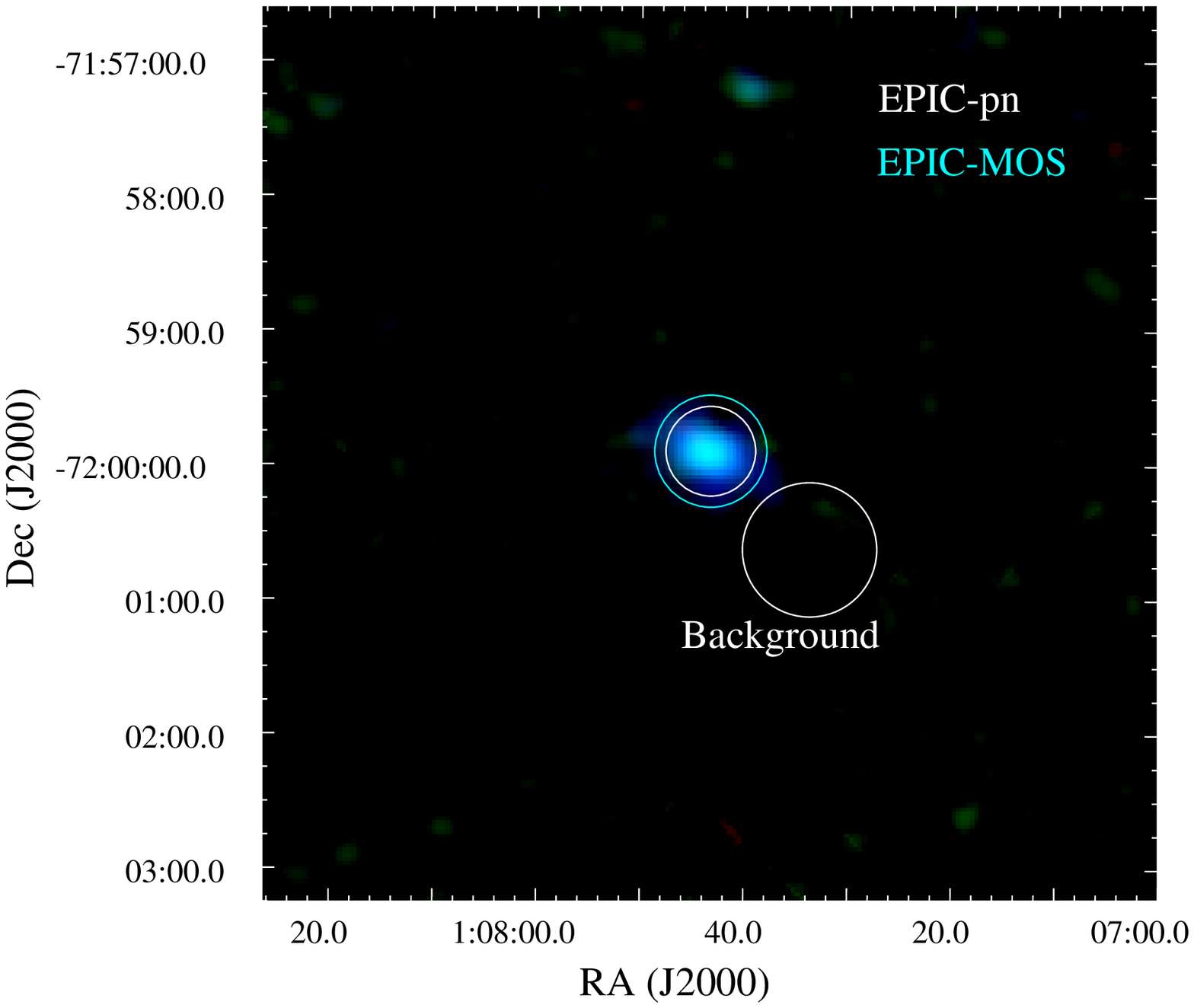}\includegraphics[angle=0,clip=]{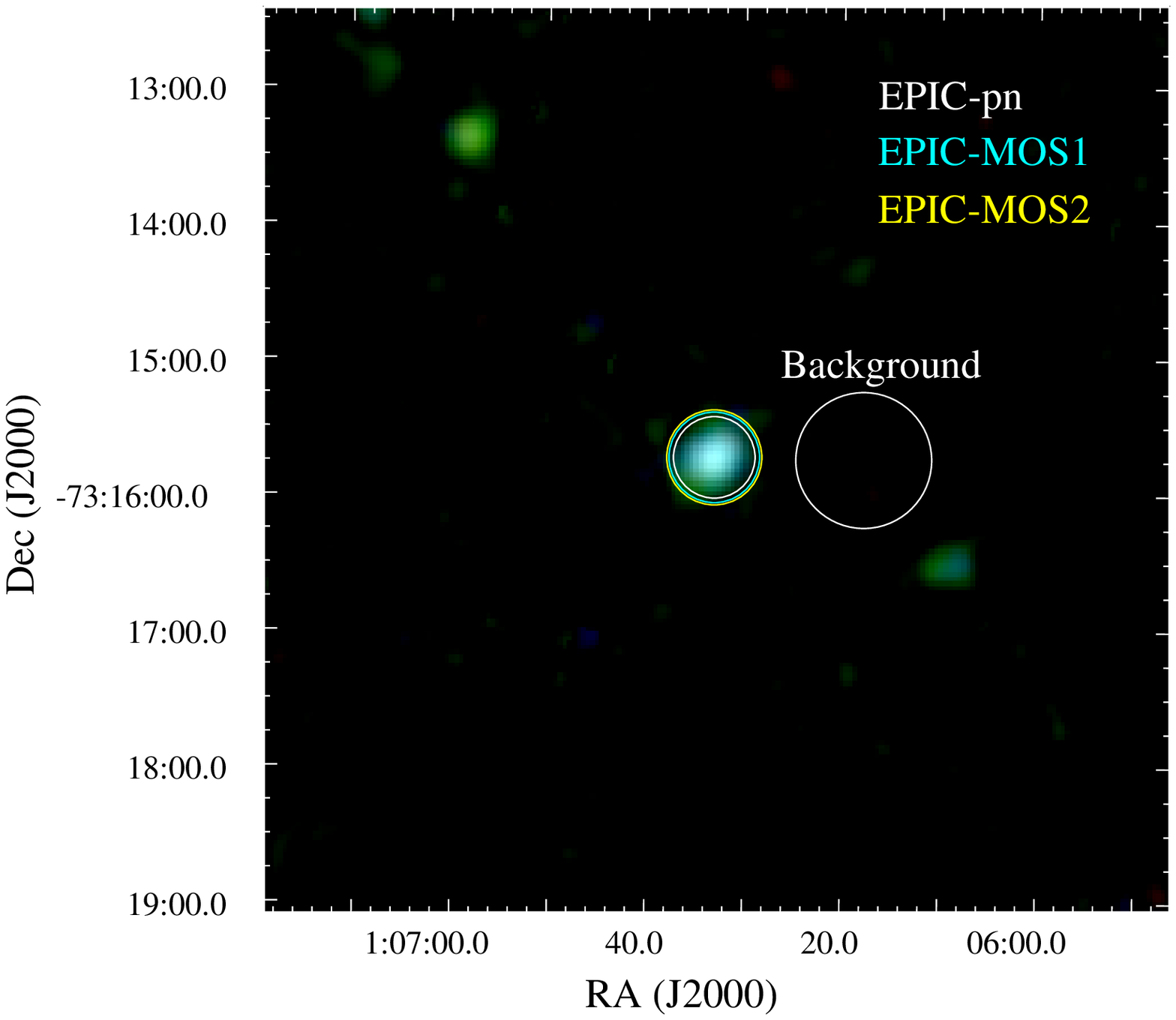}}
  \caption{ \xmm\ EPIC colour image of \xmmb\ ({\it Left}) and \xmma\ ({\it Right}). Red/green/blue show logarithmically-spaced intensities in the (0.2 -- 1.0)/(1.0 -- 2.0)/(2.0 -- 4.5) keV band. Circles indicate the extraction regions for sources and background events.
          }
  \label{fig:xima}
\end{figure*}

One of the aims of the XMM-Newton
large program SMC survey (Haberl \& Pietsch 2008a) is the ongoing
study of the Be/X-ray binary population of the SMC,
which can be used as a star formation tracer for $\sim$50
Myr old populations (Antoniou et al. 2010). Two previous systems have recently been identified from this survey:  XMMU J005011.2-730026 = SXP 214 (Coe et al., 2011) and XMMU J004814.0-732204 = SXP11.87 (Sturm et al 2011). In this paper we
present the analysis of X-ray and optical data from two newly
discovered X-ray sources \xmma ~and \xmmb. Both of these sources exhibit most of the characteristics of Be/X-ray binary systems, but only marginal evidence for X-ray pulsations from the spinning neutron star in one of the systems. If a weak, or undetectable pulsed signal is a common phenomenon, then this may have significant implications for the number of such systems in the Magellanic Clouds and the Milky Way, most of which have been primarily identified from their X-ray pulsations.

\section{X-ray Observations}

\subsection{Observations and data reduction}
The observation setup can be found in Table~\ref{tab:xray-obs}. The two X-ray sources were discovered in observations 19 and 27, respectively, of the XMM-Newton large program survey of the SMC (Haberl et al. 2012, in preparation).
\xmm\ has three X-ray telescopes (Aschenbach et al. 2002), one equipped with EPIC-pn (Str{\"u}uder et al. 2001) and two with EPIC-MOS (Turner et al. 2001) CCD detectors in the focal plane.
Both objects were covered with all three instruments.
We used XMM-Newton SAS 11.0.0\footnote{Science Analysis Software (SAS), http://xmm.vilspa.esa.es/sas/} to process the data.
Temporal screening was not necessary, since soft proton flares were at a low level (i.e. $<$8 cts ks$^{-1}$ arcmin$^{-2}$ for EPIC-pn).

We extracted single and double pixel events from EPIC-pn
and single to quadruple events from EPIC-MOS, both having {\tt flag=0}.
Background events were selected from a point source free area on the same CCD for all three instruments from a circle of radius 30\arcsec.
The source selection region was a circle with radius determined by {\tt eregionalayse} to optimise the signal-to-noise ratio.
X-ray colour images indicating the extraction regions are plotted in Fig.~\ref{fig:xima}.
Spectra and response matrices were created with {\tt especget}.
We binned the spectra to  a signal-to-noise ratio of 5 for each bin.
For time series, the photon arrival times were recalculated for the solar barycentre.
To increase the statistics for timing analysis, we also created merged time series from all three instruments in various energy bands.
Light curves, binned to have 25 cts bin$^{-1}$ on average, were created using the SAS tasks {\tt etimeget}
and corrected for background, vignetting, and point spread function losses by {\tt epiclccorr}.

\begin{table*}
  \caption{\xmm\ observations.}
  \begin{center}
    \begin{tabular}{lclclcccccr}
      \hline\hline\noalign{\smallskip}
      \multicolumn{1}{c}{ObsID} &
      \multicolumn{1}{c}{Revolution} &
      \multicolumn{1}{c}{Date} &
      \multicolumn{1}{c}{Time} &
      \multicolumn{1}{c}{Instrument} &
      \multicolumn{1}{c}{Filter} &
      \multicolumn{1}{c}{Offax$^{(a)}$} &
      \multicolumn{1}{l}{Net Exp.} &
      \multicolumn{1}{r}{Net Cts.$^{(b)}$} &
      \multicolumn{1}{r}{R$_{\rm sc}^{(c)}$}\\
      \multicolumn{1}{l}{} &
      \multicolumn{1}{l}{} &
      \multicolumn{1}{l}{} &
      \multicolumn{1}{c}{(UT)} &
      \multicolumn{1}{l}{} &
      \multicolumn{1}{l}{} &
      \multicolumn{1}{c}{[\arcmin]} &
      \multicolumn{1}{c}{[ks]} &
      \multicolumn{1}{l}{} &
      \multicolumn{1}{r}{[\arcsec]}\\

      \noalign{\smallskip}\hline\noalign{\smallskip}
      0601211901 & 1827  &    2009-11-30    &  15:08--23:37 & EPIC-pn   & thin   & 11.6 &   25.7     &   817   & 20   \\
      \xmmb           &       &                  &  14:45--23:37 & EPIC-MOS1 & medium & 10.9 &   31.1     &   405   & 25   \\
                 &       &                  &  14:45--23:37 & EPIC-MOS2 & medium & 11.6 &   31.1     &   400   & 25   \\
      0601212701 & 1840  &    2009-12-26    &  07:47--17:39 & EPIC-pn   & thin   & 7.5  &   30.2     &   529   & 18   \\
       \xmma          &       &                  &  07:24--17:39 & EPIC-MOS1 & medium & 8.5  &   36.1     &   210   & 20   \\
                 &       &                  &  07:24--17:39 & EPIC-MOS2 & medium & 8.5  &   36.1     &   228   & 21   \\
       \noalign{\smallskip}\hline\noalign{\smallskip}
     \end{tabular}
  \label{tab:xray-obs}
  \end{center}

  $^{(a)}$ Off axis angle of the sources.
  $^{(b)}$ Net counts in the $(0.2-10.0)$~keV band.
  $^{(c)}$ Radius of the source extraction region.
\end{table*}

\subsection{X-ray coordinates}
Using {\tt emldetect} for source detection, we determined positions for the X-ray sources in both observations.
For observation 0601212701 we found ten sources, correlating with Chandra X-ray sources from McGowan et al. (2008), which we use as astrometric reference.
Using this correlation, we derived an error weighted average offset of $\Delta$RA=2.71\arcsec and $\Delta$Dec=-0.47\arcsec.
The corrected best-fit position for the transient is RA(J2000)=01:06:33.10  and Dec(J2000)=-73:15:43.1.
The 1$\sigma$ position uncertainty is 0.57\arcsec, where we assume a systematic uncertainty of 0.5\arcsec (Pietsch, Freyberg \& Haberl 2005), quadratically added to the statistical uncertainty.
We assign the name XMMU\,J010633.1-731543 to this source.
Analogous for observation 0601211901, we found 25 sources detected with Chandra, allowing to derive a boresight correction of
 $\Delta$RA=-0.71\arcsec and $\Delta$Dec=1.84\arcsec, which results in a source position of RA(J2000)=01:07:43.13 and Dec(J2000)=-71:59:53.5 for XMMU\,J010743.1-715953.
The total 1$\sigma$ position uncertainty is 0.55\arcsec.

\subsection{Spectral analysis}
\begin{figure*}
  \resizebox{\hsize}{!}{\includegraphics[angle=-90,clip=]{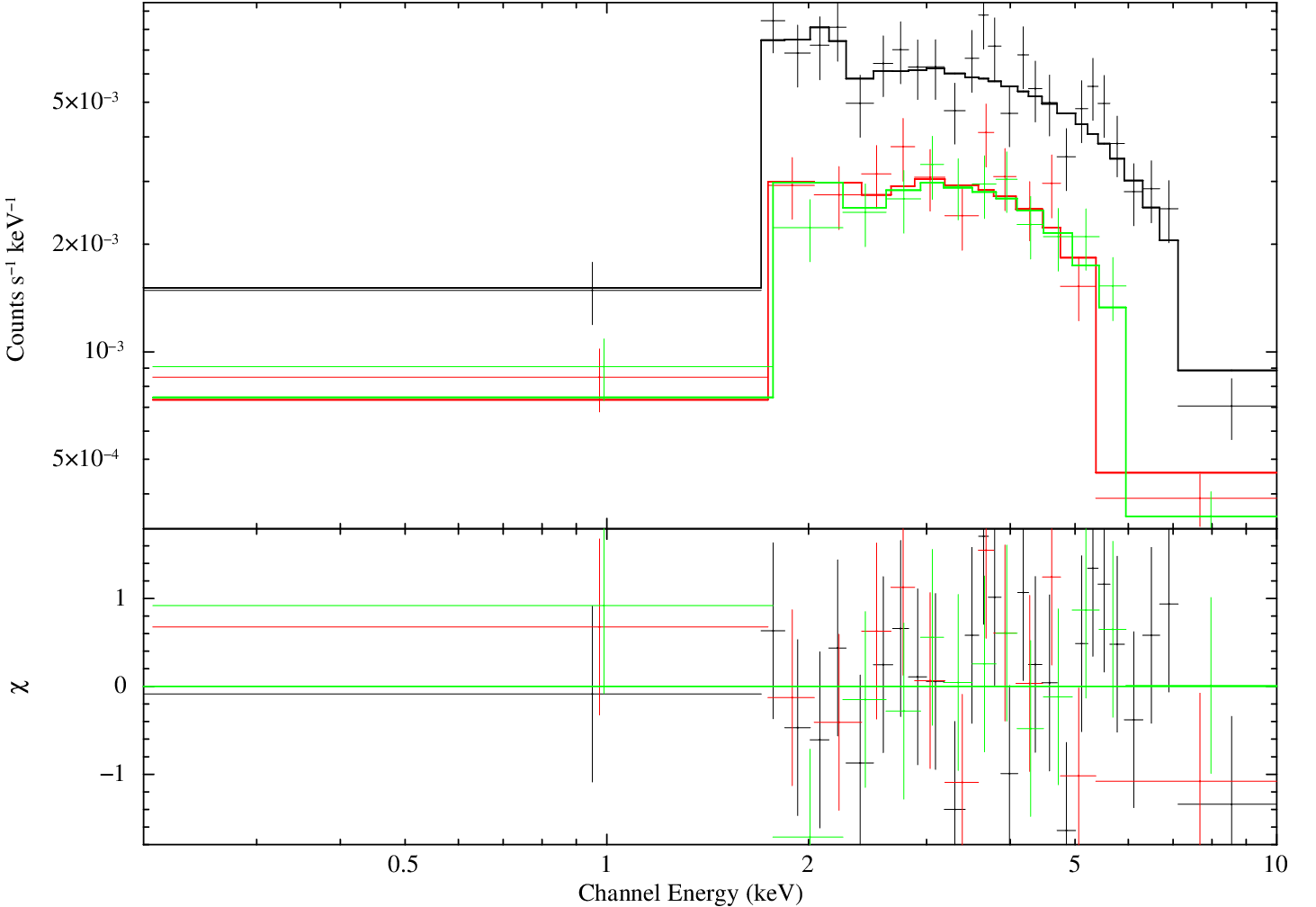}
                        \includegraphics[angle=-90,clip=]{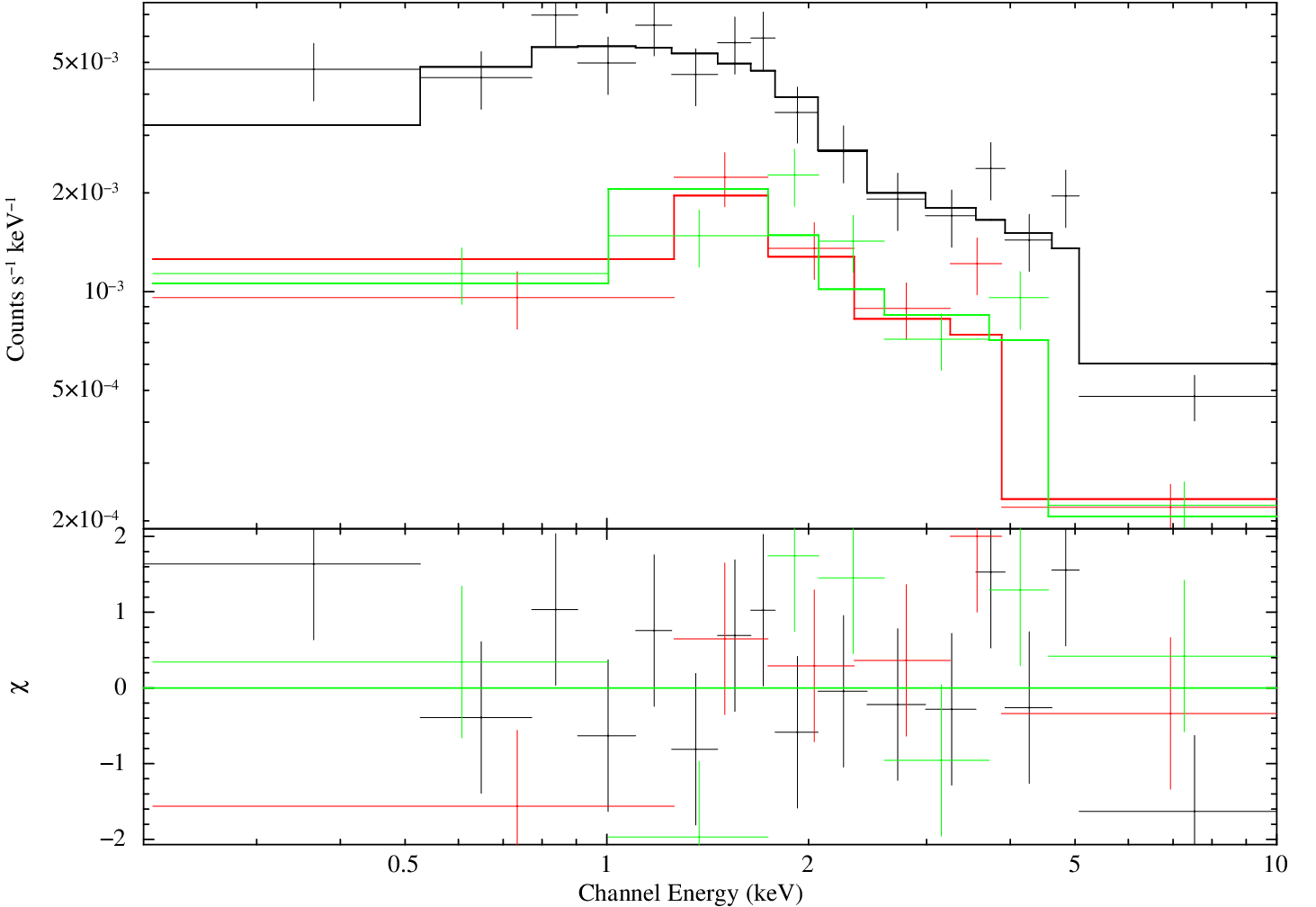}
                       }
  \caption{
           EPIC-pn (black), EPIC-MOS1 (red), and EPIC-MOS2 (green) spectra of \xmmb\ (left) and \xmma\ (right).
           Top panels show the spectra, together with the best-fit model, lower panels show the residua.
          }
  \label{fig:spectrum}
\end{figure*}

The X-ray spectra were modelled with an absorbed power-law using {\tt xspec} (Arnaud 1996) version 12.7.0.
The Galactic foreground column density was set to N$_{\rm H, gal} =  6\times 10^{20}$ cm$^{-2}$ (Dickey \& Lockman 1990) with solar abundances according to Wilms, Allen \& McCray (2000).
An additional column density N$_{\rm H, smc}$, having SMC-typical abundances set to 0.2 for elements heavier than helium (Russell \& Dopita, 1992), was a free parameter in the fit
and accounts for photo-electric absorption by the SMC interstellar medium and source intrinsic absorption.
For each source, the three EPIC spectra were fitted simultaneously.
Constant factors ($C_{\rm MOS1}$ and $C_{\rm MOS2}$) were allowed to vary to account for instrumental differences, relative to EPIC-pn ($C_{\rm pn}=1$).
The best-fit parameters are listed in Table~\ref{tab:spec}. Uncertainties and limits are given for 90\% confidence.
The spectra together with the best-fit model are plotted in Fig.~\ref{fig:spectrum}.

\begin{table*}
\caption[]{Spectral fitting results.}
\begin{center}
\begin{tabular}{lcccccccrr}
\hline\hline\noalign{\smallskip}
\multicolumn{1}{c}{source} &
\multicolumn{1}{c}{N$_{\rm H, gal}^{(a)}$} &
\multicolumn{1}{c}{N$_{\rm H, smc}^{(a)}$} &
\multicolumn{1}{c}{$\Gamma^{(b)}$} &
\multicolumn{1}{c}{C$_{\rm MOS1}$} &
\multicolumn{1}{c}{C$_{\rm MOS2}$} &
\multicolumn{1}{c}{$F^{(c)}$} &
\multicolumn{1}{c}{$L^{(d)}$} &
\multicolumn{1}{c}{dof} &
\multicolumn{1}{c}{$\chi_{\rm red}^2$} \\

\noalign{\smallskip}\hline\noalign{\smallskip} \vspace{.7mm}
\xmmb    &   0.6  &   $80.8_{-1.6}^{+2.0}$   &  $0.96_{-0.17}^{+0.19}$  &  $1.17_{-0.12}^{+0.13}$   &   $1.18_{-0.12}^{+0.13}$  & $10.1_{-5.1}^{+1.6}$  &    5.65   &   48     & 0.76 \\
\xmma    &   0.6  &   $<0.6$              &  $0.75_{-0.09}^{+0.09}$  &  $1.10_{-0.16}^{+0.18}$   &   $1.10_{-0.16}^{+0.17}$  & $ 2.3_{-0.2}^{+0.3}$  &    0.99   &   24     & 1.42 \\
\noalign{\smallskip}\hline
\end{tabular}
\end{center}
  $^{(a)}$ Column density of photo-electric absorption by the Galaxy (fixed) and the SMC (free) in $10^{21}$ cm$^{-2}$.
  $^{(b)}$ Photon index of the power law.
  $^{(c)}$ Detected flux in the (0.2--10.0) keV band in $10^{-13}$ erg cm$^{-2}$ s$^{-1}$.
  $^{(d)}$ Unabsorbed luminosity in the (0.2--10.0) keV band in $10^{35}$ erg s$^{-1}$, assuming a distance of 60 kpc.
\label{tab:spec}
\end{table*}

\subsection{Search for pulsations}

\begin{figure}
  \resizebox{\hsize}{!}{\includegraphics[angle=-90,clip=]{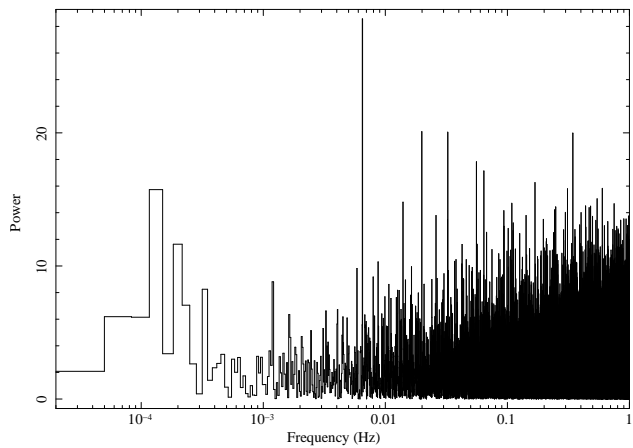}}
  \caption{
           Power density spectrum of source \xmmb\ for the merged time series in the (0.2--10.0) keV band.
          }
  \label{fig:pds}
\end{figure}

\begin{figure}
  \resizebox{\hsize}{!}{\includegraphics[angle=0,clip=]{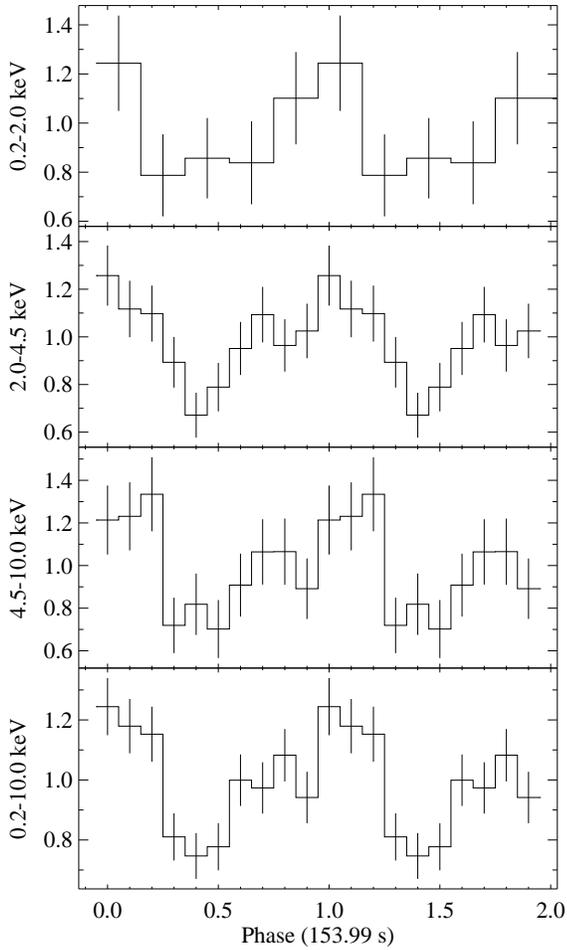}}
  \caption{X-ray pulse profile of \xmmb\ in various energy bands from the merged EPIC time series.
           The pulse profiles are background subtracted and normalized to the average net count rate of
           (0.8, 2.8, 1.8, and 5.3) $\times 10^{-2}$ cts s$^{-1}$.
          }
  \label{fig:x_pp}
\end{figure}

\begin{figure}
  \resizebox{\hsize}{!}{\includegraphics[angle=0,clip=]{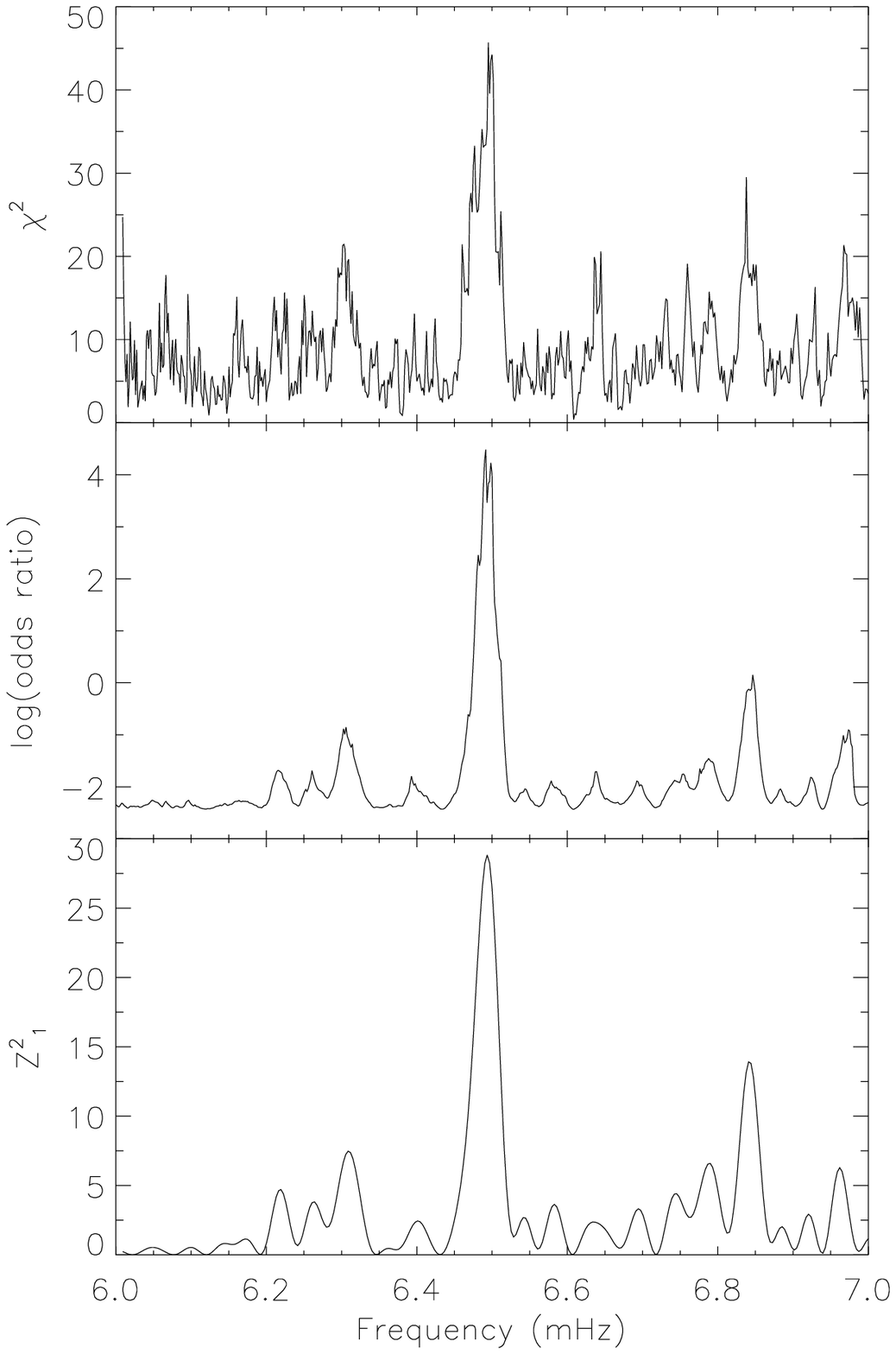}}
  \caption{
           {\it Top:} $\chi^2$ test for constancy of the light curve of \xmmb\, for trial frequencies between 6 and 7 mHz.
           {\it Middle:} Frequency dependence of the Bayesian odds ratio.
           {\it Bottom:} Rayleigh $Z^2$ test with one involved harmonic.
          }
  \label{fig:period_sta}
\end{figure}

In the case of \xmmb, we found an indication for X-ray pulsations at $6.49 \times 10^{-3}$ Hz in a Fast Fourier Transformation (FFT), as shown in Fig.~\ref{fig:pds}.
By using a Bayesian periodic signal detection method (Gregory \& Loredo 1996),
we determined the pulse period with 1$\sigma$ uncertainty to (153.99$\pm$0.13) s.
The background subtracted light curve, convolved with this period, is shown in Fig.~\ref{fig:x_pp}.
A $\chi^2$ test resulted in a maximum of $\chi^2 = 45.7$.
The significance of the signal was estimated to be 2.0$\sigma$ with a Rayleigh Z$^2_1$ test for one harmonic (Haberl \& Zavlin 2002, Bucheri et al., 1983).This assumes a number of 93000 independent trials in a frequency range of 10$^{-4}$ to 3 Hz and a
maximum Z$^2_1$ of 29.
For the statistical tests, see Fig.~\ref{fig:period_sta}.
We estimate the pulsed fraction to (18.7$\pm$4.5)\% by assuming a sinusoidal pulse profile.
Further confirmation of this pulse period is necessary, demanding higher statistics.

In the case of \xmma, we could not find any indication for pulsations in FFTs.
To estimate upper limits for the pulsed fraction, we created pulse profiles in the (0.2--10.0) keV band of the merged instrument data.
We folded the light curve at the highest peaks in the according FFT at $9.7\times 10^{-4}$, $4.0\times 10^{-3}$, $1.2\times 10^{-2}$, $8.5\times 10^{-2}$, and  $3.1\times 10^{-1}$ Hz.
Analogous to above, we fitted a sine function and determined pulsed fractions of 3\%, 21\%, 23\%, 19\%, and 25\% with typical uncertainties of $\pm5\%$ taking into account background and using 90\% confidence intervals.
Therefore we estimate the pulsed fraction of \xmma\ to be below $\sim$30\% for periods between 0.2 and 10$^4$s.

\subsection{Short term X-ray variability}

\begin{figure*}
  \resizebox{\hsize}{!}{\includegraphics[angle=-90,clip=]{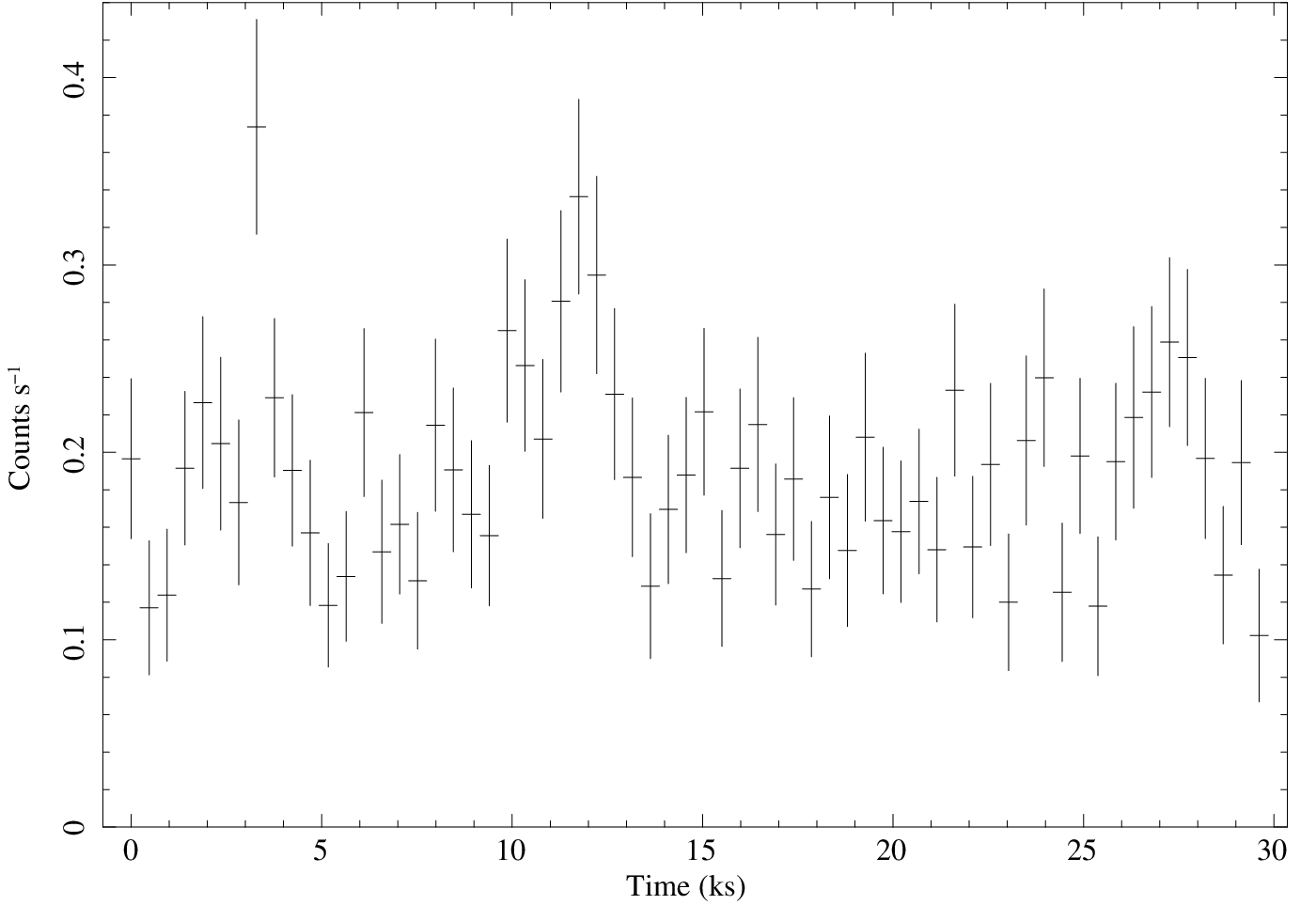}
                        \includegraphics[angle=-90,clip=]{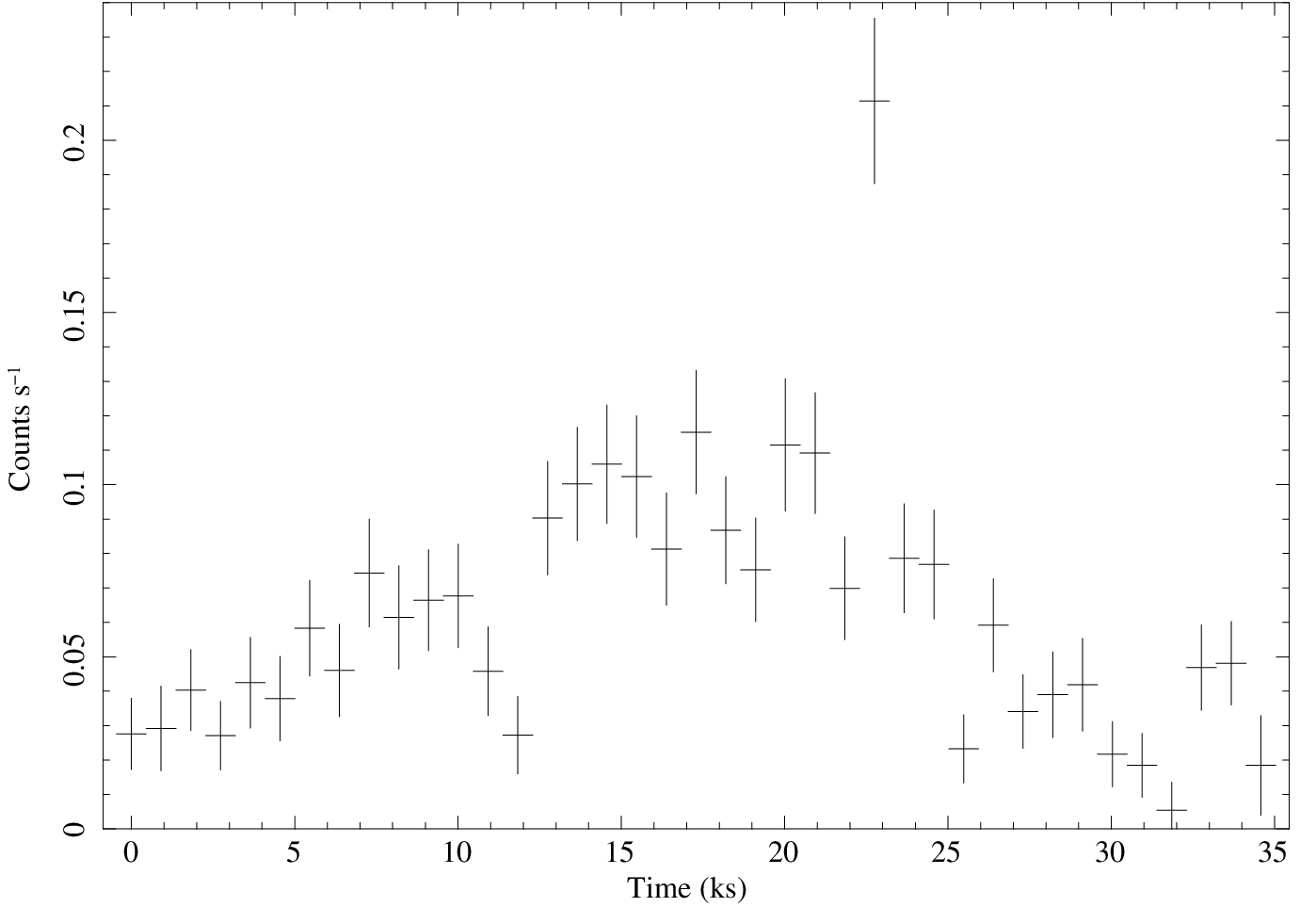}
                       }
  \caption{
           Merged corrected light curves from EPIC (0.2 -- 10 keV) for \xmmb\ (left, MJD 55165) and \xmma\ (right, MJD 55191), binned with 470 s and 910 s.
          }
  \label{fig:x_lightcurve}
\end{figure*}

The background-corrected instrument-merged light curves are shown in Fig.~\ref{fig:x_lightcurve}.
Evident is the variability of source \xmma, but also the light curve of \xmmb\ contains some indication of flare-like features.
A $\chi^2$ test for variability of the binned light curve resulted in $\chi^2/{\rm dof}$ = 101/63  corresponding to a probability of constancy of 0.16\% for \xmmb\ and
$\chi^2/{\rm dof}$ = 274/38 corresponding to a probability of $2\times10^{-36}$ for \xmma.

\subsection{Long term X-ray variability}

Source \xmmb\ was observed 3 times with Swift on 2009-12-04, 2011-05-17, and 2011-07-02.
We extracted spectra within a source region of 30\arcsec\ radius and from a background region with 200\arcsec\ radius.
Due to short exposures of 3.5, 4.3, and 2.6 ks the spectra contain only 33, 10, and 22 net counts.
Using C statistics, we fitted the same spectral model as above, only allowing the flux (an overall normalisation) to vary.
We derived fluxes of $1.3_{-0.5}^{+0.5}$, $3.1_{-1.5}^{+1.0}$, and $1.2_{-0.4}^{+2.7} \times 10^{-12}$ erg cm$^{-2}$ s$^{-1}$, consistent (within the errors) with the flux given in Table 2 for this source.
We note, that 91\%, 60\% and 82\% of the net counts are detected above 2.0 keV, pointing to a generally hard spectrum of the source
as seen during the \xmm\ observation.

There is no ROSAT source detected within 40\arcsec\ (Haberl et al., 2000).
In a 20.8 ks ROSAT PSPC observation (ObsID: rp500250n00) from 1993-10-13 to 1993-10-28 the source position was observed $\sim$15\arcmin\ offaxis.
We derive a 3$\sigma$ upper limit of 32 net cts from a circle of 60\arcsec ~radius around the source and a background region of 90\arcsec.
This translates into a source flux of $1.1 \times 10^{-12}$ erg cm$^{-2}$ s$^{-1}$
in the (0.2 -- 10.0) keV band, by assuming the same spectral model as above
and using PIMMS\footnote{Portable Interactive Multi-Mission Simulator (PIMMS), http://heasarc.nasa.gov/Tools/w3pimms.html}.
Therefore the ROSAT non-detection is also consistent with a persistent source.

The position of source \xmma\ was in the FoV of one Chandra/ACIS-I observation
of the SMC wing survey (PI: Coe, ObsID: 5496, MJD 53797).
Using the CIAO Version 4.0\footnote{Chandra Interactive Analysis of Observations (CIAO), http://cxc.harvard.edu/ciao/} tool {\tt aprates},
we derived a 3$\sigma$ upper limit of $1.44 \times 10^{-3}$ cts s$^{-1}$ in the (0.2 - 10.0) keV band.
Assuming a time-independent spectrum and using PIMMS,
this translates into a flux of $3.33 \times 10^{-14}$ erg cm$^{-2}$ s$^{-1}$. Interestingly, the Chandra non-detection falls at a time of optical brightness (see Section 3.1.1. below).
A significant source variability by a factor of at least 6.9 proves the transient behaviour of this system.

 Similarily, for \xmma, no ROSAT match was found within 40\arcsec and no upper limit deeper than the Chandra value could be derived.

\section{The optical counterparts}

\subsection{\xmma}

\begin{figure}
\includegraphics[angle=-0,width=80mm]{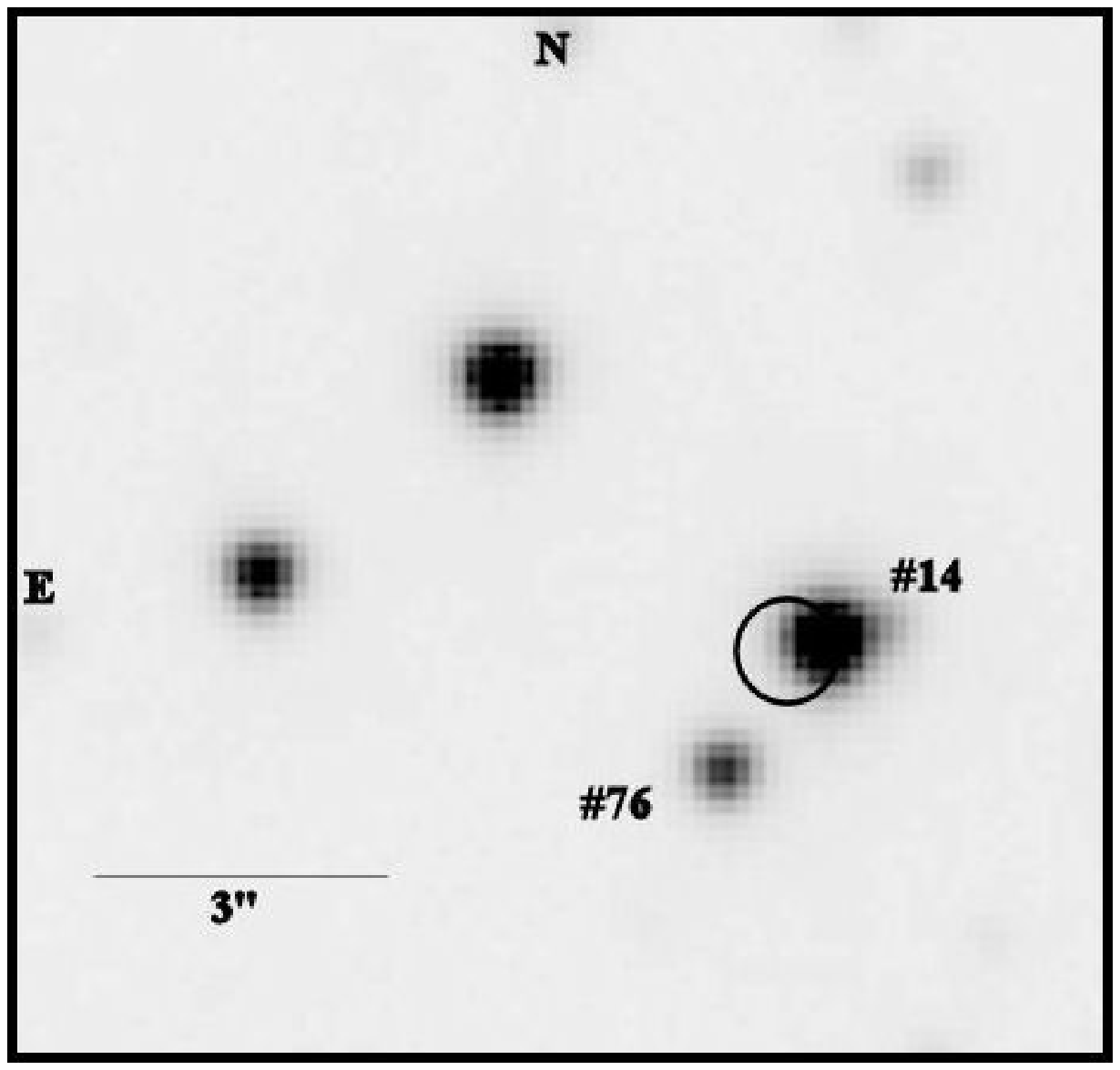}
\caption{OGLE I-band finding chart for \xmma. The circle indicates the XMM-Newton positional uncertainty. The two candidate OGLE objects are indicated as numbers 14 and
76.}
\label{fig:trans1FC}
\end{figure}

Figure~\ref{fig:trans1FC} shows the position of the XMM-Newton source located upon the OGLE I band image of the region. Unlike the image obtained from the DSSII project, this image clearly resolves the two candidates labeled as Objects 14 \& 76 (following the fuller OGLE III naming convention of SMC111.3.14 and SMC111.3.76).

\subsubsection{Optical \& IR photometry}
\begin{figure}
\includegraphics[angle=-0,width=80mm]{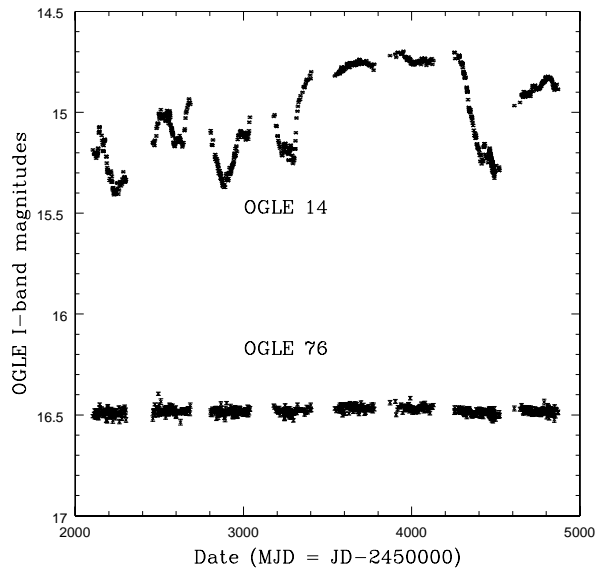}
\caption{OGLE I-band photometry for the two candidate OGLE objects for \xmma ~shown in Figure~\ref{fig:trans1FC}.}
\label{fig:figog}
\end{figure}

Figure~\ref{fig:figog} shows the OGLE-III photometry for the probable (Object 14) and possible (Object 76) candidate sources to \xmma ~is illustrated in Figure~\ref{fig:trans1FC}. It is immediately apparent that the two objects exhibit very different styles of behaviour, and also that one is always much fainter than the other. The behaviour of Object 14 is characteristic of the long-term behaviour patterns of Be stars (see Mennickent et al 2002) unlike that of Object 76. Both of these light curves were subjected to a Lomb-Scargle period analysis.

The light curve of Object 14 is extremely complex.
It displays a Type-1/Type-2 light curve, according to
the classification scheme of Mennickent et al. (2002), i.e. it shows
both outbursts and high/low states. Such composite light curves have
only been observed for 2.6\% of the stars studied by those authors.
Over a period of 3000 days the maximum amplitude is $\simeq$0.7mag
in $I$, which is high compared to most Be stars studied in
the SMC. The colour and magnitude ranges for Type-1/Type-2 objects,
according to Mennickent et al. (2002) are $15.44<V<17.01$
$-0.19<B-V<-0.00$ and $-0.19<V-I<0.23$. So, the most probable
optical counterpart of XMMU J010633.1-731543 is brighter in $V$ (see Table~\ref{tab:optir}) than most objects of similar type tabulated by Mennickent et al.

Many different attempts to detrend the data were used but none revealed any evidence for coherent behaviour indicative of either binary motion nor Non Radial Pulsations (see Bird, Coe \& McBride 2011 for a review of such characteristics in Be star companions to HMXBs).There is some evidence in Years 5 \& 6 of different behaviour, but, again, no periodic signal was seen.

On the other hand, the light curve of Object 76 is very stable and did reveal some evidence of power at a period of 15.97d. Folding the data at this period revealed a weak modulation with a possible sinusoidal structure. This kind of modulation is often associated with pulsations in stars, and the true pulse period may not be 15.97d, but rather this may be a harmonic of the true period beating with the daily sampling period.

Thus on both the XMM positional location of \xmma ~and on the evidence of the variability it seems most likely that Object 14 is the correct counterpart to the X-ray source. This object is listed as number 64194 in the catalogue of Massey (2002). In that catalogue the optical magnitudes are listed as B=14.61 and V=14.60 but there may be some danger of contamination from Object 76 if the two objects were not fully resolved. However, from our OGLE III data we have an average V-I=-0.11 which, when combined with the Massey V value would imply I=14.71. As can be seen from Figure~\ref{fig:figog} this corresponds to the brightest I band state seen by OGLE III. The Massey data were collected on JD2451187 which, unfortunately, pre-dates OGLE III and this object is not in the part of the SMC covered by OGLE II. So it is not possible to verify that the measurements presented in Massey (2002) do not represent some combination of the signals from both Objects 14 \& 76. However, the colours of (B-V)=0.01 and V-I=-0.11 would be typical for the Be stars seen in HMXBs in the SMC (Coe et al, 2005).

Object 14 also falls in the IR Sirius catalogue of the SMC (Kato et al, 2007). From there it is possible to obtain the IR magnitudes obtained on JD 2452869 which give an IR colour of J-K=0.14 - again consistent with the IR colours of HMXBs (Coe et al, 2005). All these optical and IR magnitudes are tabulated in Table~\ref{tab:optir}.

\begin{table}
\caption{I, J, H \& K photometric measurements of the Object 14 candidate for \xmma ~taken on MJD 52869. The B \& V magnitudes were taken on MJD 51187 (Massey 2002).}
\label{tab:optir}
 \begin{tabular}{ccc}
  \hline
   Waveband & Magnitude &Error on Magnitude   \\
  \hline
B & 14.61 & 0.01 \\
V & 14.60 & 0.01 \\
I & 15.33 & 0.01 \\
J & 15.52 & 0.02 \\
H & 15.46 & 0.02 \\
K & 15.38 & 0.04 \\
 \hline
 \end{tabular}
\end{table}

The Massey (2002) counterpart, object number 64194, with RA=01:06:32.9,
Dec=-73:15:42.2 (2000), has
$V=14.6\pm0.02$, $B-V=0.01\pm0.02$, $U-B=-0.92\pm0.01$,
$V-R=0.13\pm0.04$. In the MCPS catalogue of Zaritsky et al. (2002),
the closest counterpart has RA=01:06:32.98, Dec=-73:15:42.5 (2000),
which coincides with the Massey object. This star has the following
magnitudes and colors in the MCPS catalogue: $U=13.78\pm0.034$,
$B=14.681\pm0.035$, $V=15.055\pm0.037$, $I=15.072\pm0.038$ and from
2-MASS, $J=15.151\pm0.059$, $K'=14.698\pm0.10$ (as given in the MCPS
catalogue).
Thus the reddening free Q-parameter is
$Q_{UBV}=(U-B)-0.72(B-V)=-0.93$, corresponding to a B1 star using
the Massey colors, and $Q_{UBV}=(U-B)-0.72(B-V)=-0.63$ using the
MCPS magnitudes, corresponding to a B3-4 star.
All these spectral types are consistent with a Be star.

In a 5 arcmin region around this object, the average reddening given by
Zaritsky et al. (2002) is $<A_V>=0.26\pm0.28$ i.e.
$E(V-I)=0.104\pm0.112$. On the other hand, from the Haschke et al.
(2011) reddening maps, $E(V-I)=0.052\pm0.015$ (from 5 determinations
in regions centered within 5 arcmin of the object) and $A_V=
3.1\times0.052 = 0.16$.

\subsubsection{Optical spectroscopy}

Spectroscopic observations of both sources discussed in this paper were made using telescopes both at the South African Astronomical Observatory (SAAO) on 2011 Sep 22, and at the European Southern Observatory (ESO) on 2011 Dec 10. At SAAO the 1.9m telescope of the South African Astronomical Observatory (SAAO) was used solely for red spectra. A 1200 lines per
mm reflection grating blazed at 6800\AA ~was used with the SITe
CCD which is effectively 266×1798 pixels in size, creating a
wavelength coverage of 6200\AA~ to 6900\AA~. The intrinsic resolution
in this mode was 0.42\AA~/pixel. Data were reduced using Figaro \footnote{http://www.starlink.rl.ac.uk/docs/sun86.htx/sun86.html}.

 At ESO red and blue spectra were taken with the ESO Faint Object Spectrograph (EFOSC2) mounted at the Nasmyth B focus of the 3.6m New Technology Telescope (NTT), La Silla, Chile. The EFOSC2 detector (CCD\#40) is a  Loral/Lesser, Thinned, AR coated, UV flooded, MPP chip with 2048$\times$2048 pixels corresponding to 4.1\arcmin$\times$4.1\arcmin on the sky. The instrument was in longslit mode with a slit width of 1.5\arcsec. Grisms 14 and 20 were used for blue and red end spectroscopy respectively. Grism 14 has a a grating of 600~lines~mm$^{-1}$ and a wavelength range of $\lambda\lambda3095$--$5085$~\AA{} ~producing a dispersion of 1~\AA{}~pixel$^{-1}$. The resulting spectra have a spectral resolution of $\sim12$~\AA{}. Grism 20 is one of the two new Volume-Phase Holographic grisms recently added to EFOSC2. It has 1070 lines~pixel$^{-1}$ but a smaller wavelength range, from 6047--7147~\AA{}. ~This results in a superior dispersion of 0.55 \AA{}~pixel$^{-1}$ and produces a spectral resolution for our red end spectra of $\sim6$~\AA{}.~Filter OG530 was used to block second order effects.
The data were reduced using the standard packages available in the Image Reduction and Analysis Facility (\textsf{IRAF}). Wavelength calibration was implemented using comparison spectra of Helium and Argon lamps taken through out the observing run with the same instrument configuration. The spectra were normalized to remove the continuum and a redshift correction was applied corresponding to the recession velocity of the SMC.

\begin{figure}
\includegraphics[angle=-0,width=80mm]{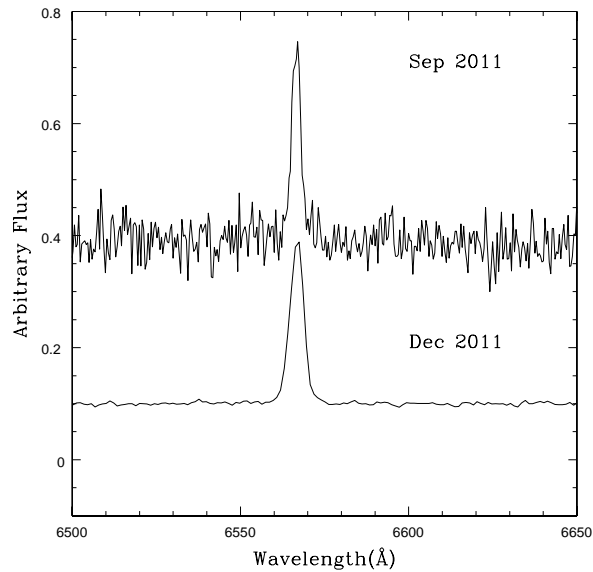}
\caption{Red end spectra of the optical counterpart to \xmma. The measured equivalent widths of the H$\alpha$ line are -14$\pm$1\AA (22 Sep 2011, upper curve) and -10.0$\pm$0.2\AA (10 Dec 2011, lower curve).}
\label{fig:ha0106}
\end{figure}

The red spectra are shown in Figure~\ref{fig:ha0106} and indicate only a small decline in the H$\alpha$ emission over few months from an Equivalent Width of -14$\pm$1\AA (SAAO 22 Sep 2011) to -10.0$\pm$0.2\AA (ESO 10 Dec 2011). In both cases the line profile is consistent with a single peak suggesting that we are viewing the circumstellar disk in this system face-on.

\begin{figure*}
\includegraphics[angle=90,width=170mm]{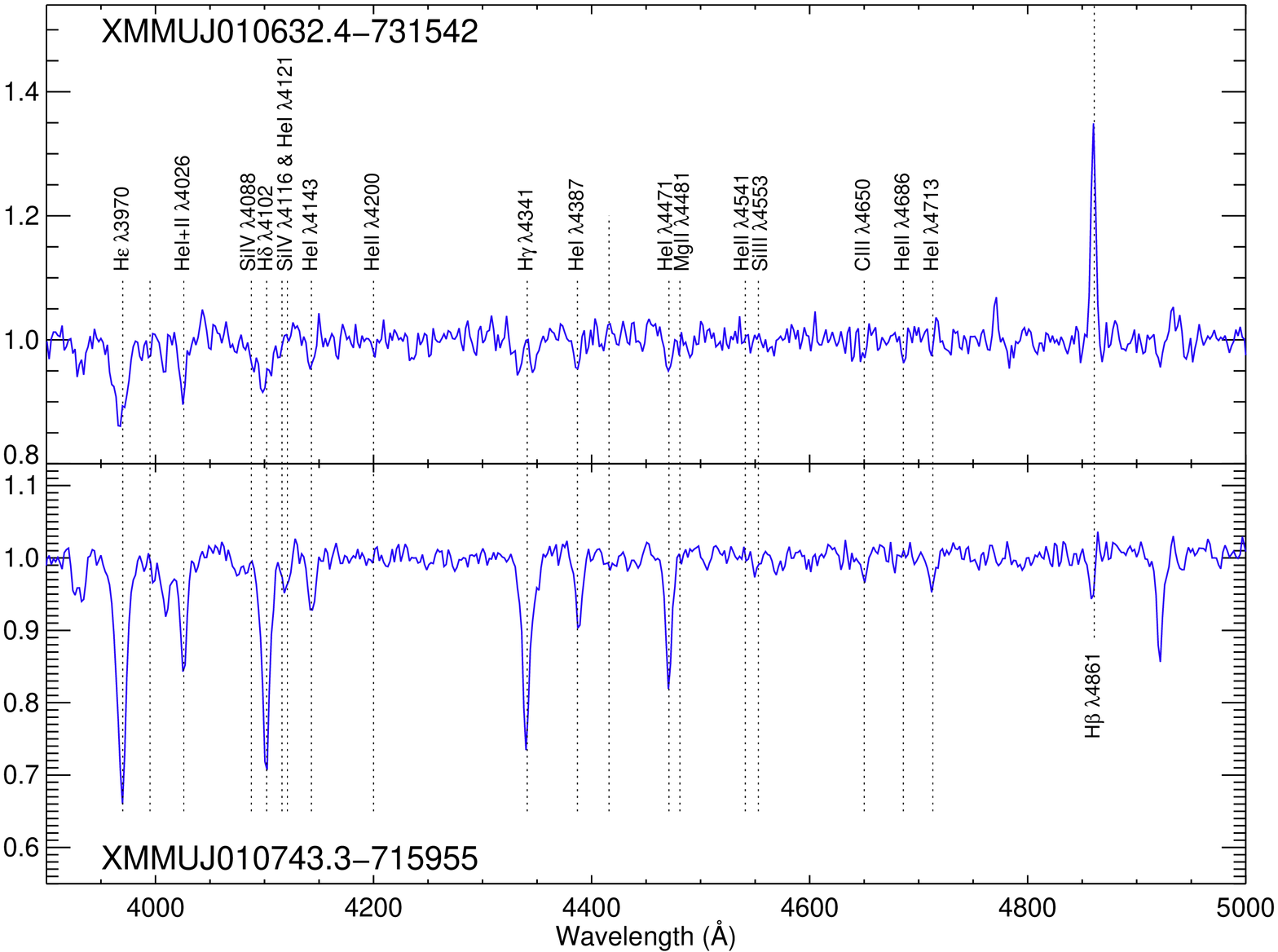}
\caption{Blue end spectra of the optical counterparts to \xmma ~and \xmmb.}
\label{fig:blue_spec}
\end{figure*}

The blue spectra of the OGLE 14 counterpart to the XMM object is shown in Figure~\ref{fig:blue_spec} and it can be used to establish the spectral classification of this object.
OB stars in our own galaxy are classified using the ratio of certain metal and helium lines (Walborn \& Fitzpatrick 1990) based on the Morgan-Keenan (MK; Morgan et al. 1943)) system. However, this is unsuitable in lower metallicity environments as the metal lines are either much weaker or not present. As such, the optical spectra of XMMUJ010632.4-731542 and XMMUJ010743.3-715955 were classified using the method developed by Lennon (1997) for B-type stars in the SMC and implemented for the SMC, LMC and Galaxy by Evans et al (2004, 2006). This system is normalized to the MK system such that stars in both systems show the same trends in their line strengths. The luminosity classification method from Walborn \& Fitzpatrick (1990) was assumed in this work.

The optical spectrum of \xmma ~shows evidence of Si\textsc{iv} lines at $\lambda4088$ and $\lambda4116$ implying the star cannot be later than type B1. The He\textsc{ii} lines are difficult to distinguish above the noise level of the spectrum, however the presence of a line at $\lambda$4686 and the apparent lack of any line at $\lambda$4541 would suggest a spectral classification of B0.5, as such we classify \xmma ~as a B0.5-1e star.
The luminosity class was determined using the ratios of S III $\lambda4553$ to H I $\lambda4387$ and He I $\lambda4143$ to He I $\lambda4121$. The latter increases dramatically with increasing luminosity class (from I-V) whereas the former decreases with luminosity class. The relative strengths of these lines suggests a luminosity class of V for \xmma. So the proposed classification for \xmma ~is B0.5-1Ve.

This spectral classification can be checked against the photometry. If we assume that the reddening and brightening is
mainly due to the contribution from the cooler equatorial disk, then
we can take the color corresponding to the faintest magnitude as
corresponding to the star, reddened basically only by interstellar
extinction. For the faintest V mag recorded, $V-I=-0.124$, giving
$(V-I)_o=(V-I)-E(V-I)=-0.124-0.052=-0.176$ which is consistent with
a B6 MS star (or using the Zaritsky reddening,
$(V-I)_o$=-0.124-0.104=-0.23 consistent with a B4-5 MS star). The V
magnitude corresponding to the bluest (low state) color is
$V=15.19$, i.e. $V_0=15.19-0.16=15.03$, and for the distance modulus of the SMC
of $18.90\pm0.18$ (Kapakos et al. 2011) $M_V=-3.87$ consistent with
a B1 MS star, where we have used the Haschke et al. reddening value.
For the Zaritsky et al. reddening value, the corresponding value is
$M_V=-3.97$ more consistent with a B0 MS star.

So both the Q-parameter and the above reasoning lead the photometric data to suggest a B0-B5
main sequence star, consistent with the spectral classification.

\subsection{\xmmb}

\begin{figure}
\includegraphics[angle=-0,width=80mm]{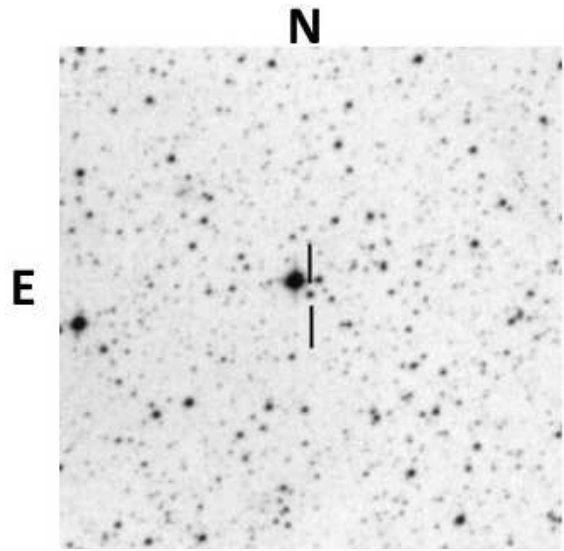}
\caption{DSSII R-band finding chart (5 x 5 arcmin) for \xmmb. The position of the optical counterpart is indicated.}
\label{fig:trans2FC}
\end{figure}

Figure~\ref{fig:trans2FC} shows the position of the XMM source located upon the DSSII R band image of the region. One object clearly lies in the XMM uncertainty region and is indicated in this figure - the OGLE III object SMC 113.4.6850. This object is also known as [MA93] 1640 (Meyssonnier \& Azzopardi, 1993). They describe the H-alpha emission of
this object as weak but sharp.
The candidate is also to be found in the 2MASS catalogue as 2MASS01074325-7159539 with  $J=16.293\pm0.101$,
$H=16.253\pm0.250$ and $K=15.839\pm0.259$.
It is also recorded in the DENIS catalogue of Cioni et
al. 2000, with $I=16.358\pm0.07$ and $J=16.591\pm0.24$.

\subsubsection{Optical \& IR photometry}

\begin{figure}
\includegraphics[angle=-0,width=80mm]{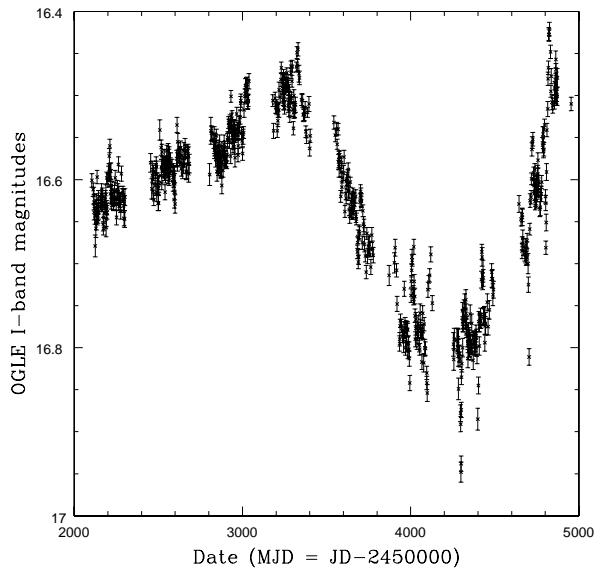}
\caption{OGLE I-band photometry for the \xmmb ~candidate OGLE object SMC 113.4.6850 shown in Figure~\ref{fig:trans2FC}.}
\label{fig:figog2}
\end{figure}

Figure~\ref{fig:figog2} shows the OGLE-III I-band photometry for the proposed candidate source to \xmmb ~illustrated in Figure~\ref{fig:trans2FC}. It is immediately apparent that there are significant brightness changes in the system typical of Be star behaviour.

The candidate for \xmmb ~also falls in the IR catalogue of the SMC (Kato et al, 2007) with the name Sirius 01074343-7159539. From there it is possible to obtain the IR magnitudes obtained on JD 2452943 (30 Oct 2003) which give an IR colour of J-K=0.40 - again very consistent with the IR colours of HMXBs (Coe et al, 2005). It is, however, rather faint with its J magnitude being the faintest of all know Be/X-ray binary systems in the SMC. The IR values were confirmed by further observations with the same camera and telescope six years later on JD 2455179 (14 Dec 2009). In addition to the IR measurements, an OGLE-III I-band measurement exists for the same date as the first IR data set. All these magnitudes are tabulated in Table~\ref{tab:optir2}.

\begin{table}
\caption{I, J, H \& K photometric measurements of the \xmmb ~candidate - see text for details.}
\label{tab:optir2}
 \begin{tabular}{ccc}
  \hline
   Waveband & Magnitude & Magnitude    \\
   & MJD52943 & MJD55179 \\
  \hline
I & 16.56$\pm$0.01 & \\
J & 16.40$\pm$0.02 &16.36$\pm$0.05\\
H & 16.21$\pm$0.02 &16.23$\pm$0.09\\
K & 16.00$\pm$0.06 & 16.03$\pm$0.22\\
 \hline
 \end{tabular}
\end{table}

\subsubsection{Optical spectroscopy}

Spectroscopic observations were taken from both SAAO (2009 Dec 14 \& 2011 Sep 22) and ESO (2011 Dec 10) as described above in Section 3.1.2.

In comparison to \xmma, ~the blue optical spectrum \xmmb ~shown in Figure~\ref{fig:blue_spec} appears featureless around the H$\delta$ line. There is no indication of any He II or Si IV lines suggesting a spectral classification later than B1.5. The presence of the Si III line at $\lambda4553$ sets a limit of B3 on the spectral classification. Stars of spectral type B2-3 are classified based on the relative strength of the Si III with respect  to the Mg II line at $\lambda4481$. It is difficult to determine if this line is even present, however we can rule out the case of a stronger Mg\textsc{ii} $\lambda4481$ with respect to Si III excluding a classification of B3. We tentatively classify XMMUJ010743.3-715955 as a B2e star. As for the previos object, the luminosity class was determined using the ratios of S III $\lambda4553$ to H I $\lambda4387$ and He I $\lambda4143$ to He I $\lambda4121$. The relative strengths of these lines suggests a luminosity class of IV - V for \xmmb. So the proposed spectral classification for \xmmb is B2IV-Ve.

\begin{figure}
\includegraphics[angle=-0,width=80mm]{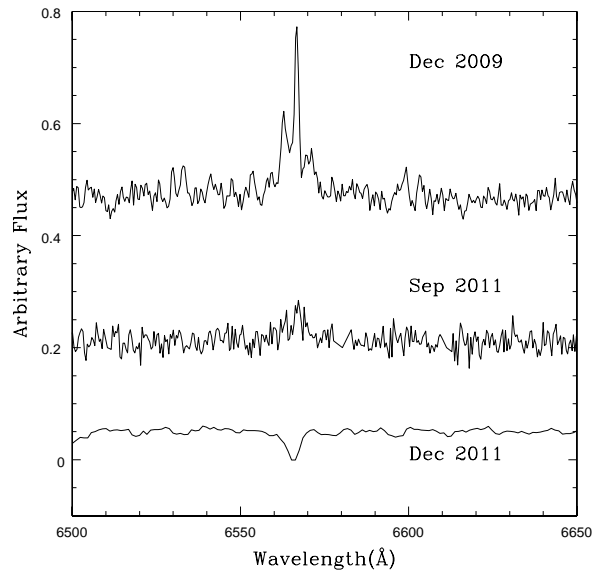}
\caption{Red end spectra of the optical counterpart to \xmmb. The measured equivalent widths of the H$\alpha$ line are -14$\pm$1\AA (14 Dec 2009), -5$\pm$5\AA (22 Sep 2011) and +1.5$\pm$0.5\AA (10 Dec 2011).}
\label{fig:ha0107}
\end{figure}

The red spectra are shown in Figure~\ref{fig:ha0107}. It is clear that H$\alpha$ line evolved from a strong, double-peaked emission feature to an absorption line showing little or no evidence for the presence of a circumstellar disk. This is supported by the lack of any H$\beta$ emission in Figure~\ref{fig:blue_spec}.

\section{Discussion and Conclusions}

\subsection{\xmma}

 The X-ray spectrum of \xmma ~shows no significant X-ray absorption in addition to the Galactic one, suggesting that this system is located on the near side of the SMC and also is probably viewed face-on. The total SMC HI column density in this direction is 3.95 $\times 10^{21}$ cm$^{-2}$.
The lack of pulsations in this object can be caused by many possible reasons e.g. weak magnetic fields, geometric effects, exceptional long pulse periods, or simply insufficient statistics on lower luminosity sources.

\begin{figure}
\includegraphics[angle=-0,width=80mm]{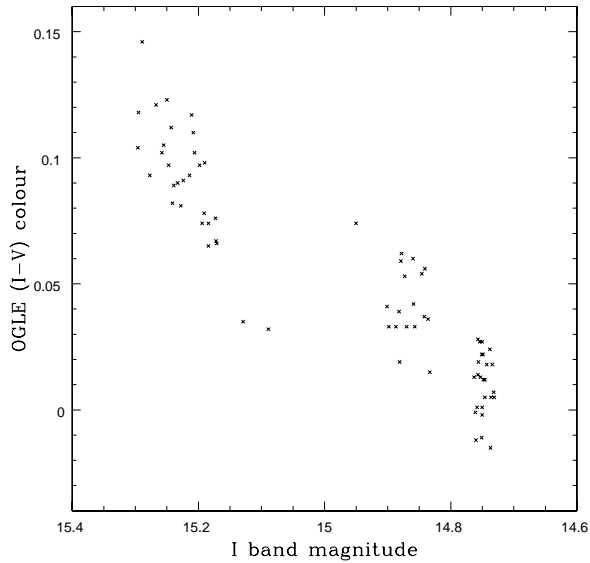}
\caption{Colour changes observed in the OGLE 14 optical counterpart to \xmma ~in the period MJD3300--4800.}
\label{fig:ogcol}
\end{figure}

It is very clear from Figure~\ref{fig:figog} that there are epochs of dramatic change in the brightness of OGLE SMC111.3.14. In particular around MJD 54200-54600 there is a considerable decrease in the I band magnitude in an linear manner on this diagram. The slope of the decrease is $\sim$0.5 mag in 130d. Similar slopes may be seen around MJD 52800-52900 and MJD 2200. In all cases the rate of change is very similar to the large change. To explore this further we used available OGLE III V-band data to study any colour changes that may be occurring. For this source there are OGLE III V-band measurements available for the period MJD53400-54800 which includes the largest variation shown in Figure~\ref{fig:figog}. Using these data it is possible to construct a magnitude-colour diagram and this is presented in Figure~\ref{fig:ogcol}. Note colour information is only available on occasions when the I and V band data were taken within 24 hours of each other. It is immediately apparent that there are substantial colour changes taking place as the source fades and brightens, in the sense that the source becomes bluer as it fades (lower right hand corner to top left corner of the figure).

Similar color-magnitude correlations in BeXRBs have also been
observed in the past, e.g. in McGowan \& Charles 2002 (MNRAS), Clark
et al.(1999), and more recently in Sarty G.E. et al.
(2011).

Janot-Pacheco et al. (1987) have shown that
changes in the emission from the circumstellar disk will produce a
correlation between V magnitude and B-V color, with the system
becoming redder as it brightens owing to the larger contribution
from the cooler disk.

An obvious interpretation for this is that the decline of the circumstellar disk inevitably takes away cooler, redder light from the overall emission of the system.

Since there are at least 3 epochs of major declines in brightness (and a couple of shorter ones) shown in Figure~\ref{fig:figog} it is obviously a phenomenon characteristic of this source. One possible explanation could be that the start of each of these declines coincides with the switching off of the stellar wind outflow, and then the slopes shown indicate the gradual dissipation of the circumstellar disk all with the same characteristic timescale. This idea would be supported by the colour changes seen at the same time.

\subsection{\xmmb}

 For source \xmmb, the total SMC HI column density is 6.3 $\times 10^{21}$ cm$^{-2}$, a factor of $\sim$12.8 smaller than the value derived from the X-ray absorption. Therefore the high X-ray absorption of \xmmb\ has to be source intrinsic. Also compared to other BeXRB, the value is quite high (Haberl, Eger \& Pietsch, 2008)and the photon distribution measured with Swift is rather hard. One possible scenario is that the neutron star is seen in/through the circumstellar disk.
Certainly the double peaked H$\alpha$ line seen in Figure~\ref{fig:ha0107}, indicates a high inclination to the system.

The Swift detection of \xmmb ~four days after the \xmm\ observation would be in agreement with a Type I outburst. In contrast to that, the two Swift detections in 2011 are separated by 46 days, and have a rather low luminosity and are therefore unlikely caused by a Type II outburst. The fact that the source is detected at 3 epochs at a rather low and similar luminosity suggest a persistent BeXRB. This would be expected, if the neutron star orbit is in the plane of the circumstellar disk of the Be star and has a low eccentricity.

The timescale of
the optical variations seen in Figure~\ref{fig:figog2} is of the order of  ten years or more. The overall
shape of the light-curve can probably be classified as a Type-3
light curve, according to Mennickent et al (2002). However it must be noted that all of
the light-curves that they classify as Type-3 have shorter
characteristic timescales of modulation (of the order of 1 yr). It
is possible (see discussion on Type-3 light-curves in their paper) that this
very long-term periodic or semi-periodic variability is linked to
global one-armed oscillations in the circumstellar envelope
(Okazaki et al., 1997). The period of the modulation is about
7 years on average and it is a weak function of spectral type,
increasing as the spectral type becomes later (see Fig.4 in Okazaki
et al 1997). The associated photometric variations are not strictly
periodic. The problem is that these oscillations are not expected
theoretically to occur in the low metallicity environment of the
SMC. However, Hummel et al. (2001) found
observational evidence contradicting this theoretical prediction.

\begin{figure}
\includegraphics[angle=-0,width=80mm]{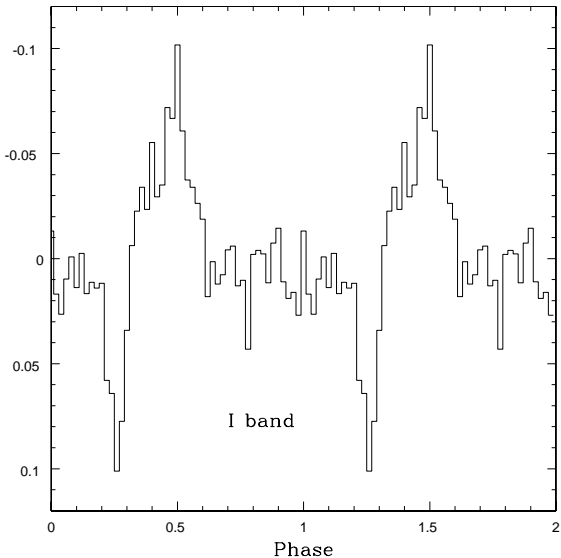}
\caption{The last three years of detrended OGLE I-band data folded at the period of 100.3d and with a $T_{o}$=MJD3872.0 for the optical counterpart to \xmmb. ~The vertical scale is detrended magnitudes.}
\label{fig:6850fold3}
\end{figure}

In addition to the general changes in flux, the presence of sharp features can be seen, in particular when the source is in a fainter state around MJD 4000-4400. If the OGLE III data are detrended and then subject to Lomb-Scargle analysis a strong peak emerges in the power spectrum at a period of 100.3d. This period is present in both the first half and the last half of the data, and not just a feature of the later years. Though it is probable that the depth of the modulation increases when the source is generally fainter. The result of folding the last three years of I-band data at the period of 100.3d is shown in the upper panel of Figure~\ref{fig:6850fold3}.

\begin{figure}
\includegraphics[angle=-0,width=80mm]{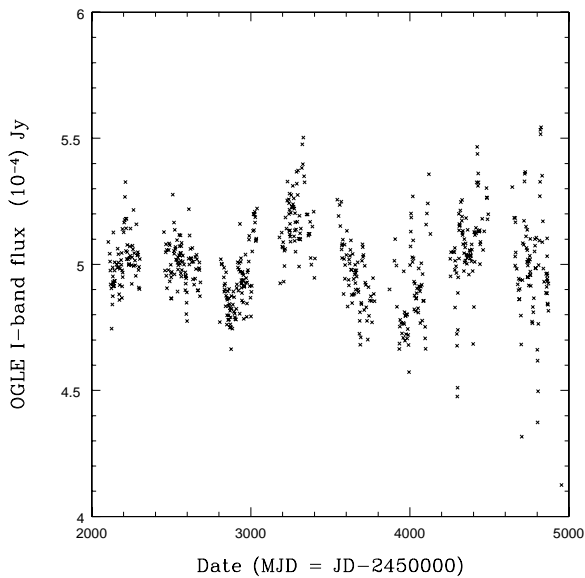}
\caption{Detrended OGLE I-band data for \xmmb ~expressed in flux units.}
\label{fig:figog3}
\end{figure}

Shown in Figure~\ref{fig:figog3} are the OGLE III I-band data converted to flux units with the global trend removed. It is apparent from this figure that the depth of the dips and spikes grows in size throughout the data set. A result with some similarities to these data has been presented by Graczyk et al, 2011 also from OGLE III data of the Magellanic Clouds. In their Figure 9 they show an example of a Transient Eclipsing Binary system in the LMC which shows strong evidence for variable-depth, eclipse-like features. These authors explain the variation in the eclipse features as due to regression of the nodes of the orbital plane and compare the result to the galactic system V907 Sco (Lacy et al, 1999). Such regression is attributed to a third body in the system.

\begin{figure}
\includegraphics[angle=0,width=80mm]{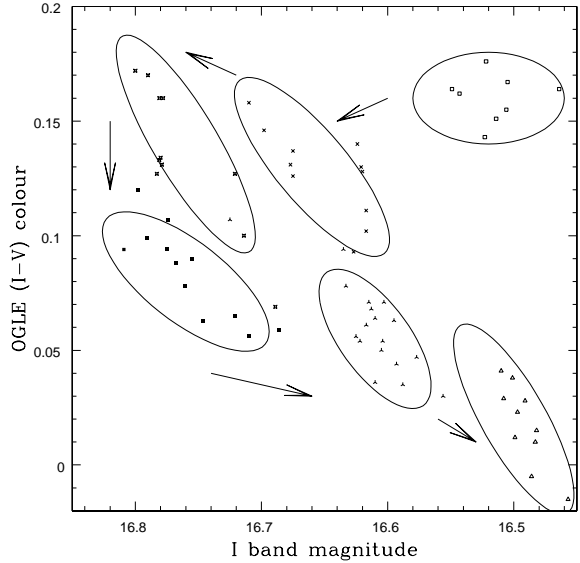}
\caption{OGLE data for \xmmb ~showing the colour change pattern. Each ellipse approximately encloses data from an annual observing season - each season has a different symbol.The total period covered by the above graph (MJD 3326 - 4953) corresponds to the major change in optical brightness in the source shown in Figure~\ref{fig:figog2}.}
\label{fig:6850ogcol}
\end{figure}

Shown in Figure~\ref{fig:6850ogcol} is the evolution of the colours of the optical counterpart to \xmmb ~seen during the major brightness change shown in Figure~\ref{fig:figog2}. From the relative positions of each year's data set it can be seen that the colour of the system follows and anti-clockwise pattern on this diagram. Similar colour evolutions seen in other SMC Be/X-ray binary systems have been presented in Rajoelimanana et al (2011). Starting in the top right hand corner of Figure~\ref{fig:6850ogcol} it can be seen that the source fades in overall brightness by about 0.3 magnitude with only small changes in colour, but then as the disk re-builds (top left to lower right) there is a significant increase in the redness of the system. This presumably reflects the growth of the disk as a large cooler component of the system. In both cases the timescales for the disk loss and subsequent rebuilding is 1-2 years.

\section{Acknowledgements}

R.S. acknowledges support from the BMWI/DLR grant FKZ 50 OR 0907.
As always, we are grateful to the support staff in SAAO for help in using the 1.9m telescope.  Also we are grateful to the IRSF/SIRIUS team, based in Nagoya
University, Kyoto University the National Astronomical Observatory of Japan, for their support prior to and during our observations. The OGLE project has received funding from the European Research Council under the European Community's Seventh Framework Programme (FP7/2007-2013)/ERC grant agreement no. 246678 to AU. We are grateful for the prompt and helpful comments from the referee.

\bsp

\label{lastpage}

\end{document}